\documentclass[%
 aip,
 amsmath,amssymb,
 reprint,%
]{revtex4-1}

\usepackage{graphicx}
\usepackage{dcolumn}
\usepackage{bm}
\usepackage{adjustbox}
\usepackage[utf8]{inputenc}
\usepackage[T1]{fontenc}
\usepackage{mathptmx}
\usepackage{xcolor}
\usepackage{hyperref}
\hypersetup{
	colorlinks   = true,
	citecolor    = blue,
	linkcolor = blue,
    urlcolor = blue
}

\begin{document}

\preprint{AIP/123-QED}

\title{A Quantum Monte Carlo study of the structural, energetic, and magnetic properties of two-dimensional (2D) H and T phase VSe$_2$}

\author{Daniel Wines}
 \email{daniel.wines@nist.gov}
\affiliation{Materials Science and Engineering
Division, National Institute of Standards and Technology (NIST),
Gaithersburg, MD 20899}

\author{Juha Tiihonen}
\affiliation{%
Department of Physics, Nanoscience Center, University of Jyv\"askyl\"a, P.O. Box 35, Finland
}%

\author{Kayahan Saritas}%

\affiliation{ 
Material Science and Technology Division, Oak Ridge National Laboratory, Oak Ridge, Tennessee 37831
}%

\author{Jaron Krogel}%

\affiliation{ 
Material Science and Technology Division, Oak Ridge National Laboratory, Oak Ridge, Tennessee 37831
}%

\author{Can Ataca}
 \email{ataca@umbc.edu}
\affiliation{%
Department of Physics, University of Maryland Baltimore County, Baltimore MD 21250
}%

\date{\today}

\begin{abstract}

Previous works have controversially claimed near-room temperature ferromagnetism in two-dimensional (2D) VSe$_2$, with conflicting results throughout the literature. These discrepancies in magnetic properties between both phases (T and H phase) of 2D VSe$_2$ are most likely due to the structural parameters being coupled to the magnetic properties. Specifically, both phases have a close lattice match and similar total energies, which makes it difficult to determine which phase is being observed experimentally. In this study, we used a combination of density functional theory (DFT), highly accurate diffusion Monte Carlo (DMC) and a surrogate Hessian line-search optimization technique to resolve the previously reported discrepancy in structural parameters and relative phase stability. With DMC accuracy, we determined the freestanding geometry of both phases and constructed a phase diagram. Our findings demonstrate the successes of the DMC method coupled with the surrogate Hessian structural optimization technique when applied to a 2D magnetic system.

\end{abstract}

\maketitle


\section{\label{sec:intro}Introduction}



\begin{figure}
\begin{center}
\includegraphics[width=8cm]{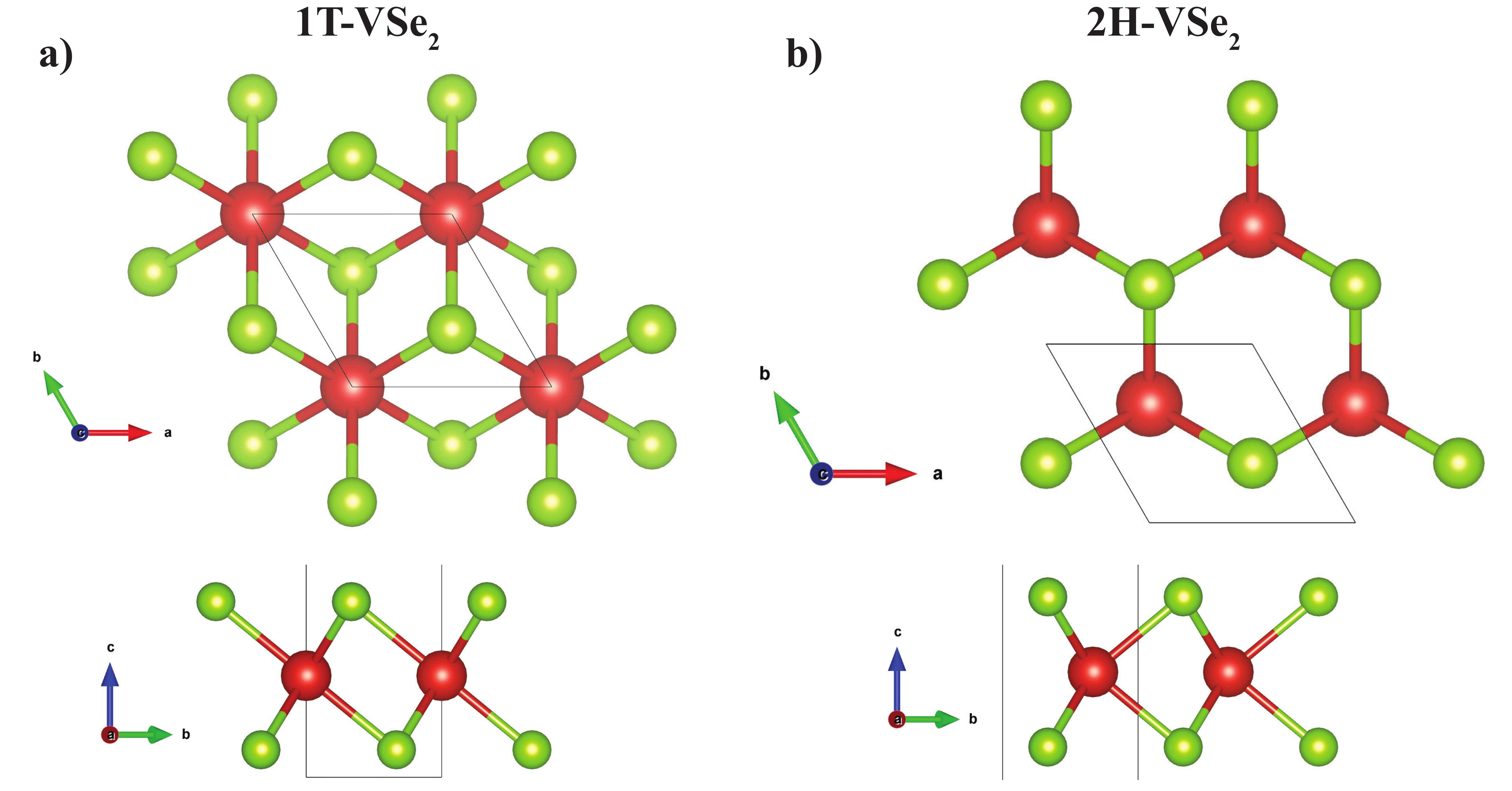}
\caption{Top and side view of the atomic structure of monolayer VSe$_2$ in the a) 1T and b) 2H phase.  }
\label{structure}
\end{center}
\end{figure}

One of the most promising two-dimensional (2D) magnetic materials that has been extensively studied experimentally and theoretically is 2D VSe$_2$. Similar to other 2D transition metal dichalcogenides (such as MoS$_2$) \cite{ataca-mx2}, VSe$_2$ exists in two phases, the T (octahedral phase (1T)-centered honeycombs) phase which is metallic and the H (the trigonal prismatic phase (2H)-hexagonal honeycombs, see Fig. \ref{structure}) phase which is semiconducting. Several experimental and theoretical studies have controversially claimed near-room temperature ferromagnetism in VSe$_2$, with conflicting results throughout the literature. Density functional theory (DFT) along with classical Monte Carlo simulations have been used to obtain an estimate of the Curie temperature of H-VSe$_2$ (291 K) \cite{vse2-bkt}, but the model Ising Hamiltonian used did not take into account the magnetic anisotropy energies, which are essential for an accurate estimation of the Curie temperature of a 2D lattice. The Curie temperature of multilayered 2D H-VSe$_2$ has been experimentally measured to be 425 K, with the ferromagnetism softening as the thickness of the sample increases \cite{vse2-exp}. Additionally, the experimental Curie temperature for monolayer T-VSe$_2$ has ranged from 300 K to 470 K \cite{vse2,https://doi.org/10.1002/adma.201903779} depending on which substrate is used (MoS$_2$, graphite, SiO$_2$-coated silicon). The experimental magnetization of T-VSe$_2$ has also been met with controversy, with values of 15 $\mu_B$ and 5 $\mu_B$ (per formula unit) being reported from two separate studies \cite{vse2,vse2-moment-exp}. Insight has also been reported with regards to how the ferromagnetism is enhanced with defects, molecular adsorption and the choice of substrate for VSe$_2$ \cite{D0NR04663A,https://doi.org/10.1002/adma.201903779,vse2}. A wide range of values have also been reported for the charge density wave (CDW) transition temperature for T-VSe$_2$, ranging from 120 K to 350 K \cite{PhysRevB.101.014514,cdw-enhanced,PhysRevLett.121.196402,vse2-exp,vse2-moment-exp}. 

These discrepancies in the electronic and magnetic properties of either phase of 2D VSe$_2$ arise from the structural parameters of each phase being coupled closely to the magnetic and electronic properties and the external factors (substrates, defects) of the individual samples. One example of this has been a reported discrepancy on which phase (T or H) is energetically more favorable. Both the T and H phases have a close lattice match and similar total energies, which makes it difficult to determine which phase is being observed experimentally. Recently, it has been reported experimentally that the T phase is favored for bulk VSe2, but with dimensionality decrease, the H phase is favored \cite{struc-phase,vse2-exp}. It has also been reported that a T-to-H phase transition can be realized by thermal annealing \cite{struc-phase}. This same structural phase transition has even been reported by applying a biaxial strain of $\approx$ 3 $\%$ (from calculated results) \cite{struc-phase,D0NR04663A,C9CP03726H}. Researchers have proposed that this lattice strain can be induced by the mismatch that occurs from putting 2D VSe$_2$ on a substrate \cite{D0NR04663A,C9CP03726H}.

From a computational perspective, results for VSe$_2$ depend heavily on which methodology is employed. In most cases, DFT with an empirical Hubbard correction (+$U$) for correlated electrons is used \cite{PhysRevB.57.1505}. For example, if the $U$ correction is applied for T and H-VSe$_2$, the T phase is more energetically favorable, while if no $U$ correction is applied, the H phase is more favorable \cite{C6CP06732H}. In addition to the discrepancies in results calculated with DFT+$U$, results between van der Waals (vdW) corrected functionals and hybrid functionals are also inconclusive \cite{C6CP06732H} in terms of predicting the relative phase stability. In order to alleviate the uncertainty in DFT methods, more sophisticated methods can be used such as Diffusion Monte Carlo (DMC) \cite{RevModPhys.73.33}. DMC is a correlated, many-body electronic structure method that has demonstrated success for the electronic and magnetic properties of a variety of bulk and 2D systems \cite{PhysRevX.4.031003,PhysRevB.94.035108,doi:10.1063/5.0023223,wines2021pathway,mno2-qmc,https://doi.org/10.48550/arxiv.2209.10379,doi:10.1063/5.0079046,PhysRevMaterials.5.024002,staros}. This method has a weaker dependence on the starting density functional and $U$ parameter and can successfully achieve results with an accuracy beyond the DFT+$U$ \cite{RevModPhys.73.33}.



Due to the fact that T and H-VSe$_2$ have structural parameters that are coupled to their electronic and magnetic properties, it makes it difficult to produce conclusive results that rely solely on DFT or DFT+$U$. For this reason, we employed our recently developed energy-based surrogate Hessian method for structural optimization with stochastic electronic structure theories (such as DMC) \cite{doi:10.1063/5.0079046} to obtain the geometry of T and H-VSe$_2$ with DMC accuracy, resulting in high-accuracy bond lengths that resolve previous functional dependent structural discrepancies. After obtaining an accurate geometry for both structures, we constructed a phase diagram between T and H-VSe$_2$ using DMC calculated energies and obtained accurate magnetic properties of each structure. The accurate estimates for lattice geometry, relative phase energy and the DMC phase diagram assist in clarifying previously inconclusive theoretical and experimental results regarding T and H
phase VSe$_2$. For full details of the computational methods used, see the Supporting Information (SI).

As an initial starting point for our study, we performed benchmarking DFT and DFT+$U$ calculations using a variety of density functionals (local density approximation (LDA)\cite{PhysRev.136.B864}, Perdew-Burke-Ernzerhof (PBE)\cite{PhysRevLett.77.3865}, and strongly constrained and appropriately normed (SCAN)\cite{PhysRevLett.115.036402} meta-GGA functionals, see SI for more details) and the Vienna Ab initio Simulation Package (VASP) code for monolayer T-VSe$_2$ and H-VSe$_2$. The goal of these simulations were to assess how sensitive the relative energy between the T and H phase is with respect to functional and material geometry. Another goal of these simulations was to benchmark the structural parameters of each material with respect to several density functionals. It is advantageous to perform these reference calculations with VASP and PAW pseudopotentials as a precursor to the more expensive DMC calculations due to the fact that they require a much smaller cutoff energy and are more cost effective for a large number of simulations. It is important to note that for all DFT and DMC simulations, we assumed a ferromagnetic ground state for both T and H-VSe$_2$. Although recent reports have suggested that T-VSe$_2$ could be experimentally paramagnetic \cite{vse2-exp}, we infer that this paramagnetism can be induced by magnetic anisotropy. In addition, the modeling of paramagnetism with computational methods imposes a great challenge, which is why we focus on the freestanding ferromagnetic ground states of both phases. A more robust treatment of the magnetic structure can be explored in future work, but is beyond the scope of this work which primarily focuses on determining the geometric structure and phase stability of 2D T and H-VSe$_2$.

\begin{figure*}
\begin{center}
\includegraphics[width=12cm]{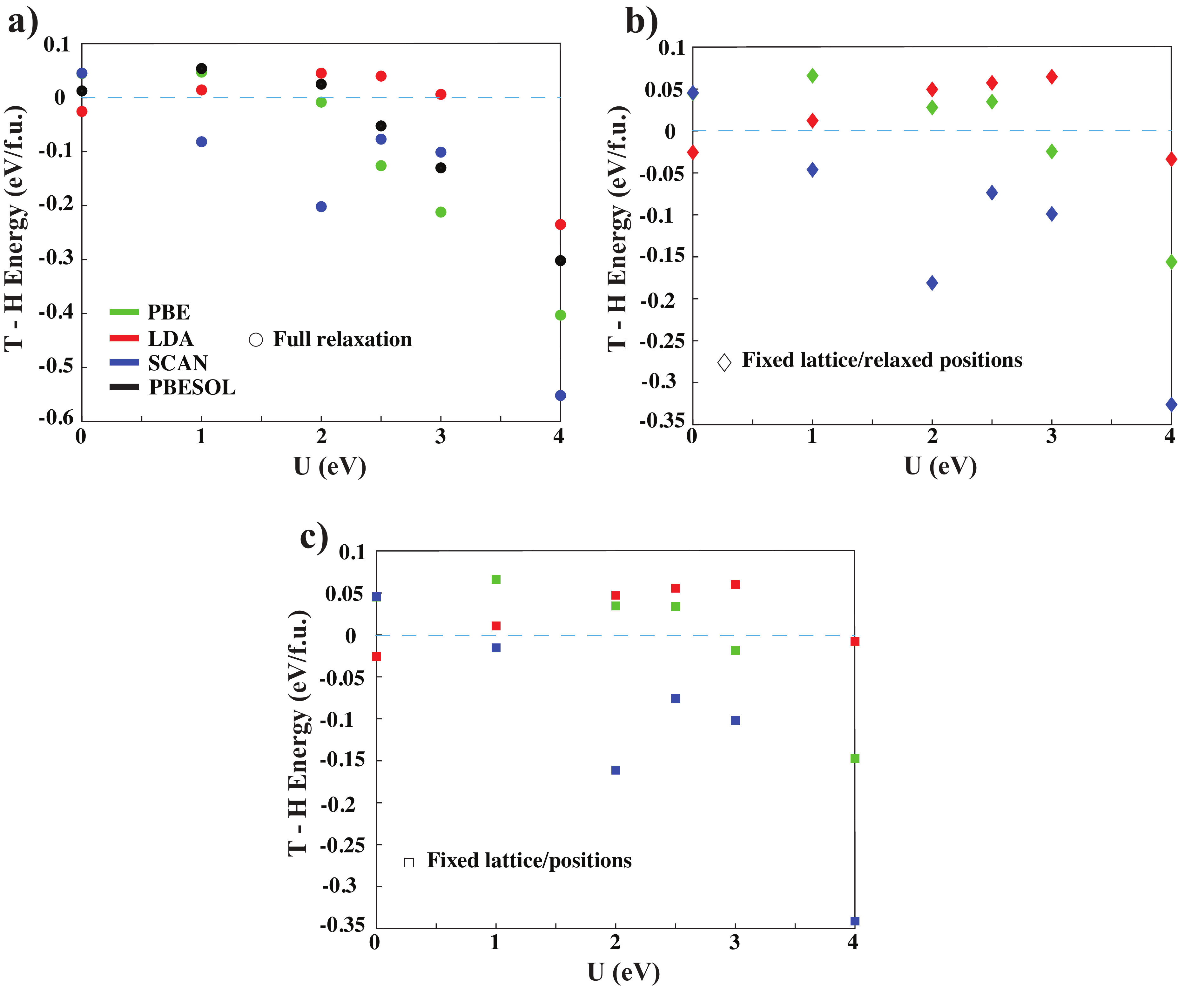}
\caption{Relative (T - H) energy between T and H phase 2D VSe$_2$ as a function of $U$ parameter for several density functionals and methods of atomic relaxation: a) fully relaxing the structure, b) fixing the lattice and atomic positions to the $U$ = 0 eV relaxed geometry of that particular functional and calculating the static energy, c) fixing the lattice to the $U$ = 0 eV relaxed geometry of that particular functional and relaxing just the atomic positions. The dotted line indicates 0 eV. }
\label{fullbench}
\end{center}
\end{figure*}

In Fig. \ref{fullbench} we present a comprehensive look at the difference in total energy between T-VSe$_2$ and H-VSe$_2$, using several DFT functionals under different geometric constraints. We performed these calculations for a variety of $U$ values in three different ways: fully relaxing the structure at each value of $U$ (Fig. \ref{fullbench} a)
), fixing the lattice and atomic positions to the $U$ = 0 eV relaxed geometry of that particular functional and calculating the static energy at each value of $U$ (Fig \ref{fullbench} b)), fixing the lattice to the $U$ = 0 eV relaxed geometry of that particular functional and relaxing just the atomic positions at each value of $U$ (Fig. \ref{fullbench} c)). The results in Fig. \ref{fullbench} indicate that there is a significant disagreement between DFT functionals, $U$ value used, and material geometries, with all three factors playing a significant role in the energy difference between T and H phase. Specifically, regardless of relaxation method, all bare (no $U$ correction) SCAN, PBE, and PBEsol functionals predict H favorable, while bare LDA predicts T favorable. For all functionals, there is a critical value of $U$ that reverses the relative phase stability, which is dependent on functional and relaxation method. The SCAN functional with a $U$ correction predicts T phase favorable, with larger energy differences. As seen in Fig. \ref{fullbench}, the trends in the relative phase stability between Fig. \ref{fullbench} b) and c) are nearly identical, but significantly vary from Fig. a). This implies that the density functional is strongly coupled to material geometry, but the lattice constant change has more of an effect on phase stability than atomic positions and bond distances. This is most prevalent for higher $U$ values ($>$ 2 eV), where the relaxed geometry changes more drastically with $U$. The interrelated nature of the material's geometry, density functional, and value of $U$ are reasons to seek out higher levels of theory beyond DFT/DFT+$U$ such as DMC to accurately determine the optimal geometry and relative energy between the phases of 2D VSe$_2$. 

\begin{figure}
\begin{center}
\includegraphics[width=8cm]{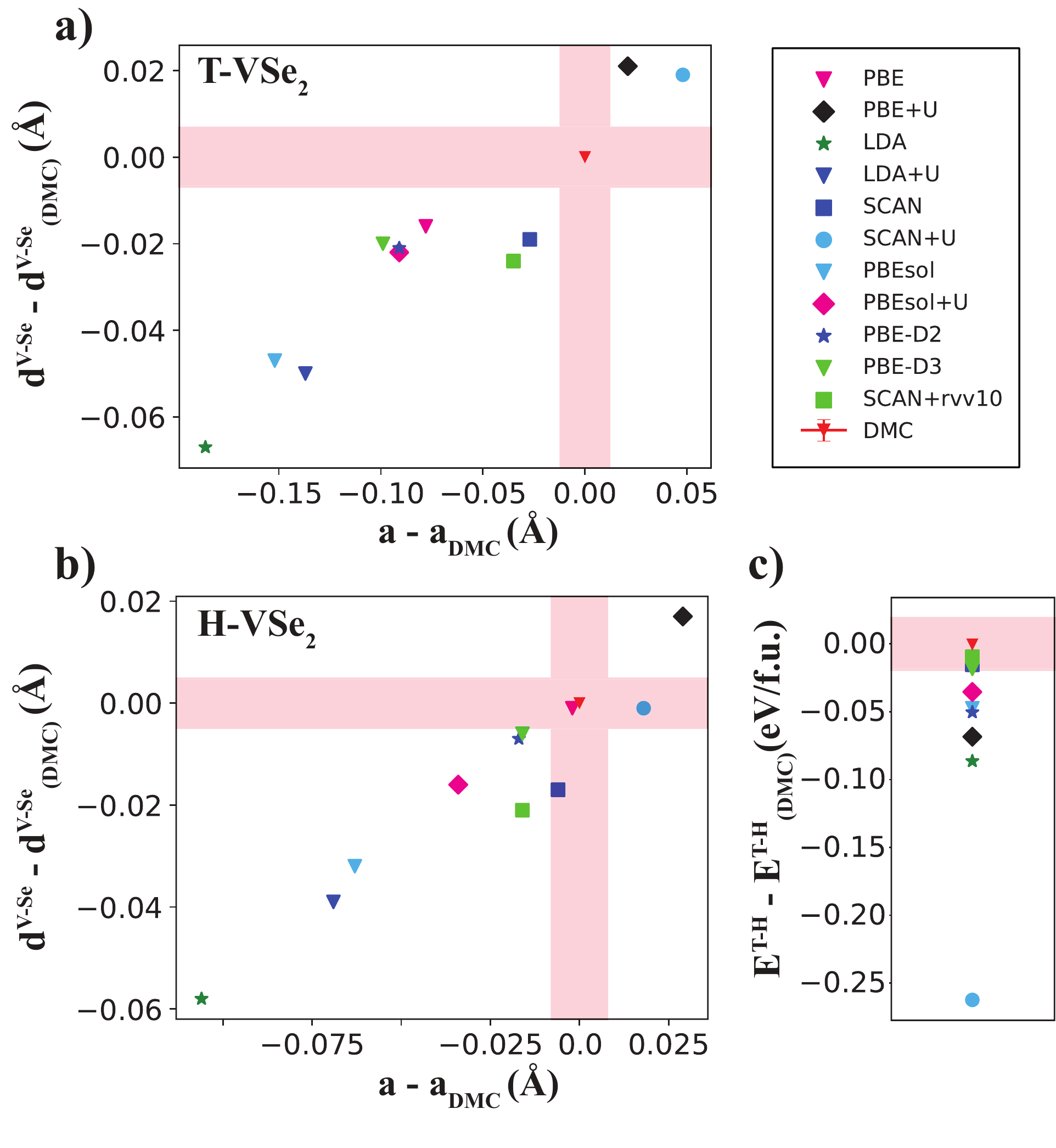}
\caption{A summary of the deviation of the geometric properties relative to the DMC calculated geometric properties for a) T-VSe$_2$ and b) H-VSe$_2$ and c) the the deviation of T - H energy relative to the DMC calculated T - H energy for a variety of DFT functionals ($U$ = 2 eV), where the DMC error bar (standard error about the mean) is represented by the red bars.}
\label{main_geofig}
\end{center}
\end{figure}

\begin{table*}[]
\caption{\label{geotable} Tabulated results for lattice constant, V-Se distance, and relative energy (T - H) for both T and H phase 2D VSe$_2$ for several computational methods. DMC error bars (standard error about the mean) are included in parenthesis.}
\begin{tabular}{l|ll|ll|l}
           & T-VSe$_2$ &      & H-VSe$_2$ &        &            \\
\hline
\hline
Method     & $a$ (\AA) & $d^{\textrm{V}-\textrm{Se}}$ (\AA) & $a$ (\AA) & $d^{\textrm{V}-\textrm{Se}}$ (\AA) & E$^{\textrm{T} - \textrm{H}}$ (eV/f.u.) \\
\hline
\hline
PBE        & 3.336                 & 2.489                      & 3.333                 & 2.502                      & 0.045                            \\
\hline
PBE+$U$=2      & 3.435                 & 2.526                      & 3.364                 & 2.520                      & -0.008                           \\
\hline
LDA        & 3.228                 & 2.438                      & 3.229                 & 2.445                      & -0.026                           \\
\hline
LDA+$U$=2      & 3.277                 & 2.455                      & 3.266                 & 2.464                      & 0.045                            \\
\hline
SCAN       & 3.387                 & 2.486                      & 3.329                 & 2.486                      & 0.045                            \\
\hline
SCAN+$U$=2     & 3.462                 & 2.524                      & 3.353                 & 2.502                      & -0.202                           \\
\hline
PBEsol     & 3.262                 & 2.458                      & 3.272                 & 2.471                      & 0.013                            \\
\hline
PBEsol+$U$=2   & 3.323                 & 2.483                      & 3.301                 & 2.487                      & 0.025                            \\
\hline
PBE-D2     & 3.323                 & 2.484                      & 3.318                 & 2.496                      & 0.010                            \\
\hline
PBE-D3     & 3.315                 & 2.485                      & 3.319                 & 2.497                      & 0.042                            \\
\hline
SCAN+rvv10 & 3.379                 & 2.481                      & 3.319                 & 2.482                      & 0.051                            \\
\hline
DMC       & 3.414(12)              & 2.505(7)      &  3.335(8)                  & 2.503(5)       & 0.06(2)     \\
\hline
\end{tabular}
\end{table*}


The relaxed lattice constants, V-Se distances, and T - H energies from Fig. \ref{fullbench} a) are presented in Table \ref{geotable} and Fig. \ref{main_geofig}, along with additional VASP reference calculations performed with the vdW corrected functionals (PBE-D2 \cite{doi:10.1002/jcc.20495}, PBE-D3 \cite{https://doi.org/10.1002/jcc.21759}, SCAN+rvv10 \cite{PhysRevX.6.041005}). The DMC computed parameters are also given for comparison in Table \ref{geotable} and Fig. \ref{main_geofig} (more discussion to follow). We observe a $\approx$ 7 $\%$ variability in lattice constant across the different methods for T-VSe$_2$ and a $\approx$ 4 $\%$ variability in lattice constant across the different methods for H-VSe$_2$. Between both phases, we observe a $\approx$ 3 $\%$ variability in V-Se distance ($d^{\textrm{V}-\textrm{Se}}$). Most strikingly, the energy difference between the T and H phases (E$^{\textrm{T} - \textrm{H}}$) drastically varies depending on the material geometry and computational methodology, ranging from -0.2 eV/f.u. to 0.06 eV/f.u.. Due to the fact that a strain-induced phase transition has been reported between T- and H-VSe$_2$ \cite{struc-phase,D0NR04663A,C9CP03726H}, we decided to perform additional VASP benchmarking calculations that involved the application of tensile and compressive strain for each monolayer. We performed these calculations for PBE, SCAN, and LDA (with $U$ = 0 eV and $U$ = 2 eV), starting from the $U$ = 0 eV geometry for each functional. The resulting equations of state are depicted in Fig. S3. As seen in the figure, the equation of state and resulting strain-induced phase transition is entirely  dependent on the functional and $U$ value, with no consistent trend. 

\begin{figure*}
\begin{center}
\includegraphics[width=12cm]{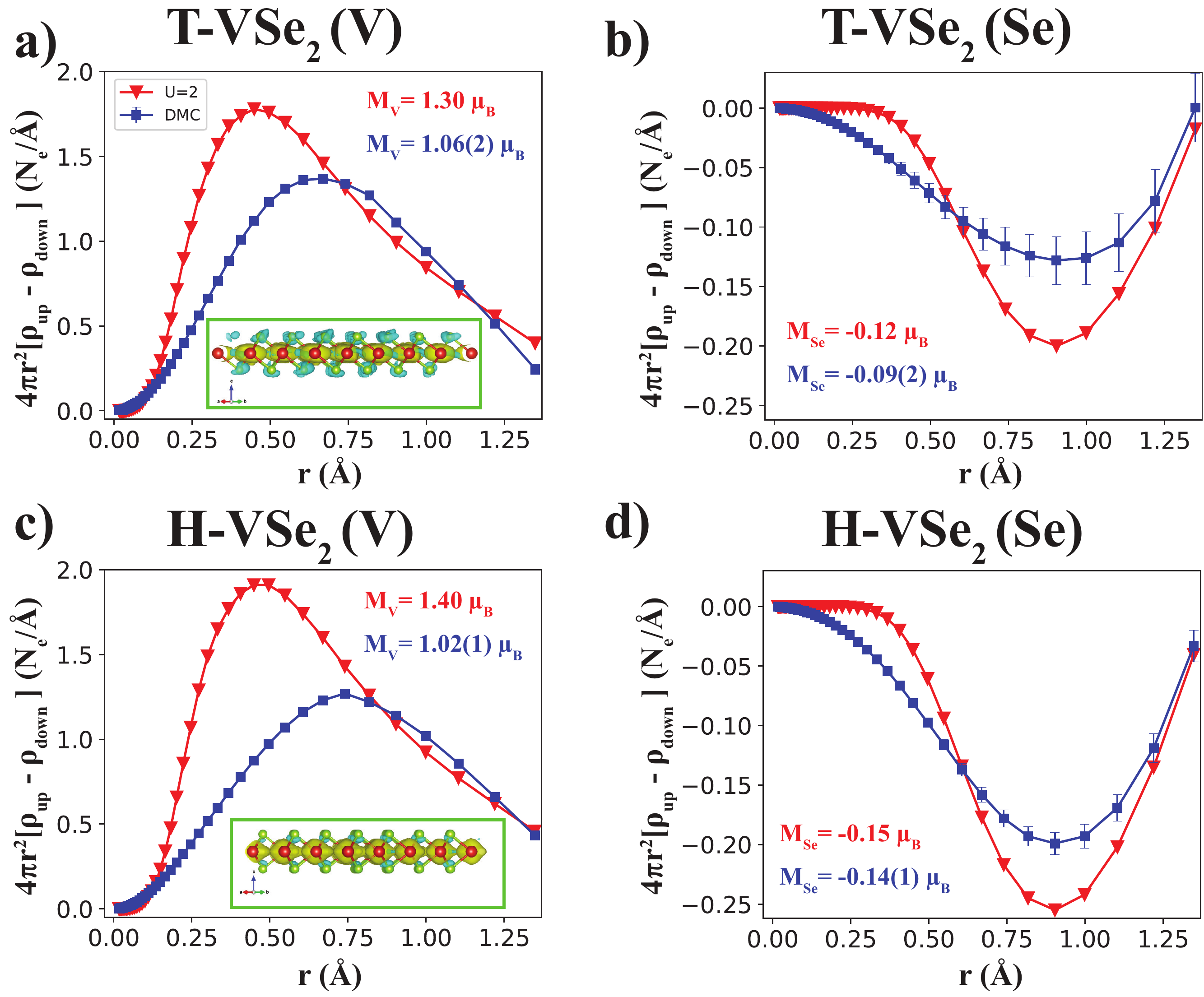}
\caption{The radially averaged spin density ($\rho_{up}$ - $\rho_{down}$) as a function of distance, calculated with DMC and PBE+$U$ ($U$ = 2 eV) of a) V and b) Se for 2D T-VSe$_2$ and c) V and d) Se for 2D H-VSe$_2$. The inset of a) and c) depicts the spin isosurface density of T-VSe$_2$ and H-VSe$_2$ respectively, where the isosurface value was set to 6 x 10$^{-3}$ e/\AA$^{3}$. The standard error about the mean for DMC is indicated by error bars in blue.}
\label{spindensity}
\end{center}
\end{figure*}

The strong sensitivity of each monolayer with respect to geometry and functional are grounds for using a higher-order method such as DMC 
to obtain a statistically accurate estimate of the lattice parameters and relative energy between phases. Prior to performing the DMC/line-search calculations, we optimized our nodal surface (orbitals selected for DFT wavefunction generation). Since DMC has the zero-variance property, it means that as the trial wave function approaches the exact ground state, 
the statistical fluctuations in the energy reduce to zero \cite{RevModPhys.73.33}. Although there have been instances where various sophisticated methods have been used to optimize the nodal surface \cite{PhysRevB.48.12037,PhysRevB.58.6800,PhysRevE.74.066701,PhysRevLett.104.193001}, we employed the PBE+$U$ approach, where the Hubbard ($U$) value was used as a variational parameter to optimize the nodal surface using DMC (similar to other successful DMC studies of magnetic materials \cite{PhysRevX.4.031003,PhysRevMaterials.5.064006,PhysRevMaterials.3.124414,PhysRevMaterials.2.085801,staros,mno2-qmc,https://doi.org/10.48550/arxiv.2209.10379}). We performed these calculations for both T and H-VSe$_2$ (24 atom supercells), where we tuned the $U$ value from (1 to 4) eV while creating the trial wavefunction and computed the DMC energy. The results of these calculations are depicted in Fig. S4, where we observe that $U$ = 2 eV yields the lowest energy for both phases. It is important to note that for the H phase, the DMC energies for $U$ = 1 and $U$ = 2 eV are statistically identical. Based on this, we created the trial wavefunction using PBE+$U$ ($U$ = 2 eV) for all subsequent DMC calculations within the surrogate Hessian line-search for both phases (all 52 DMC energy evaluations). Since we obtained an optimal $U$ value of 2 eV for both materials, we focused our DFT+U benchmarking efforts more on $U$ = 2 eV (Fig. \ref{main_geofig}, Fig \ref{spindensity}, Table \ref{geotable}, Fig. \ref{fullbench}, Fig. S3).

\begin{figure}
\begin{center}
\includegraphics[width=8cm]{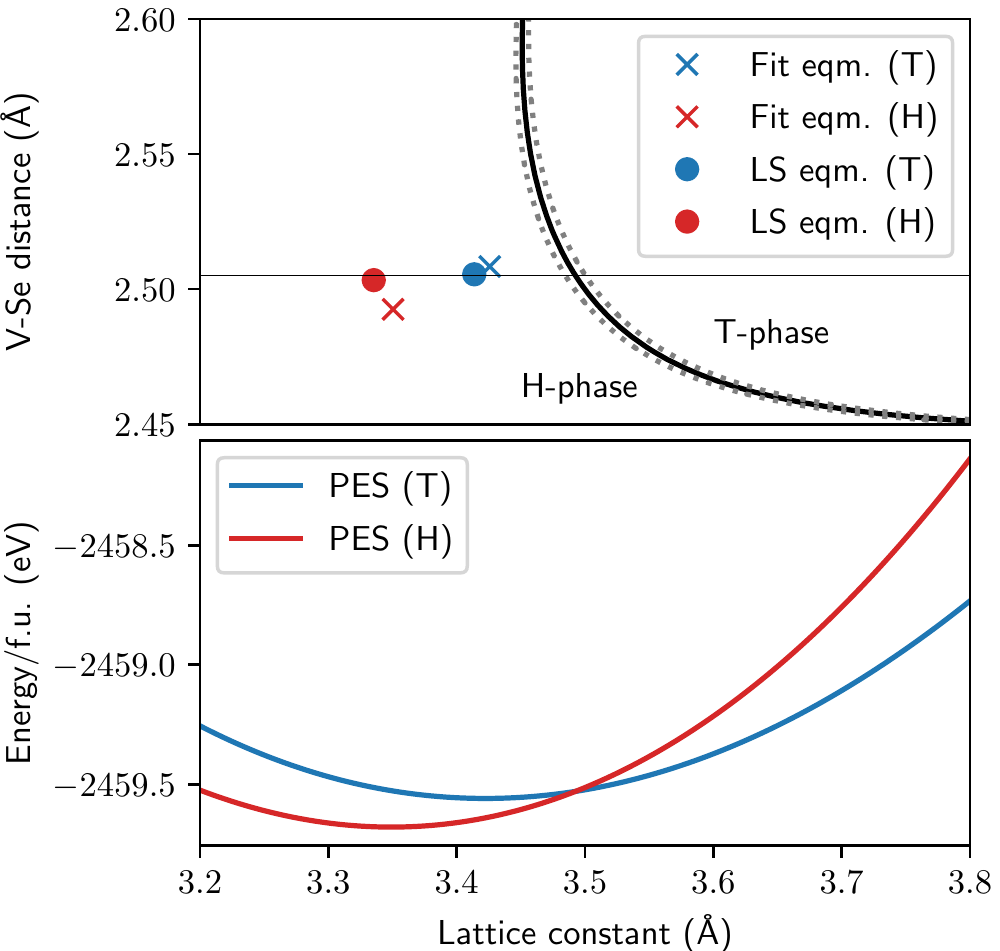}
\caption{(Top) The phase diagram of 2D VSe$_2$ in terms of $a$ and $d^{\textrm{V}-\textrm{Se}}$. The phase boundary (solid line, black) is estimated from bicubic fits. To assure quality of the fits, the estimated $\pm 0.01$ eV error contours (dotted line) and the minima from the fits ('x') and the line-search ('o') are all well separated. (Bottom) Slices of the PES at $d^{\textrm{V}-\textrm{Se}} = 2.505$ \AA.}
\label{phase_diagram}
\end{center}
\end{figure}

Based on the DMC line-search results, we determined accurate bounds on the lattice parameter ($a$) and off-plane displacement of Se ($z$), within an error tolerance of $0.018$ \AA\space or lower for both parameters. This translates to within $\approx 0.5 \%$ accuracy in a parameter set of $a$ and $d^{\textrm{V}-\textrm{Se}}$ with 95\% confidence. Convergence (absence of significant displacements outside of the error tolerance) was achieved after two parallel line-search iterations for both phases. This convergence is illustrated in Fig. S5, where the convergence of the parameter offsets of $a$ and $z$ and the convergence of the total energy per f.u. are depicted for both T and H phase 2D VSe$_2$ for the 
initial DFT relaxed structure 
(1) and both subsequent iterations of DMC (2 - 3). In addition, the final energy of both of the fitted structures (square points) are given.

The final geometric parameters and relative phase energies determined with DMC are given in Table \ref{geotable} and Fig. \ref{main_geofig}. For T-VSe$_2$, we determined a lattice constant of 3.414(12) \AA\space and a V-Se distance of 2.505(7) \AA\space. For H-VSe$_2$, we determined a lattice constant of 3.335(8) \AA\space and a V-Se distance of 2.503(5) \AA\space. The DMC finite-size extrapolated energy difference (T - H) between the two phases was determined to be 0.06(2) eV/f.u., indicating that in freestanding form at the equilibrium geometry, H-VSe$_2$ is favored over T-VSe$_2$. When comparing these DMC results to the other DFT functionals in Table \ref{geotable} and Fig. \ref{main_geofig}, it is clear that very few DFT functionals can reproduce the DMC results for lattice constant, V-Se distance and relative energy difference. The SCAN functional comes the closest to reproducing all three simultaneous DMC values, but still falls slightly short for the V-Se distances of both phases and the lattice constant of T-VSe$_2$. The fact that SCAN+U successfully predicts the structural properties (for H-VSe$_2$) and the fact that SCAN+rvv10 produces an energy difference closest to the average DMC energy difference for both phases loosely implies that a simultaneous description of correlated magnetism and vdW interactions are both needed to correctly represent the physics of VSe$_2$. Experimental measurements of the lattice constant and V-Se distance of freestanding monolayer VSe$_2$ are scarce and often times dependent on external factors such as the substrate (more discussion to follow) and sample preparation technique \cite{https://doi.org/10.1002/adma.201903779,vse2,PhysRevB.102.115149,LIU2018419}. However, Chen et al. \cite{PhysRevB.102.115149} have recently reported a lattice constant of 3.4 \AA\space for thin films of T-VSe$_2$ and Liu et al. \cite{LIU2018419} have recently reported a lattice constant of 3.3 \AA\space for epitaxially grown monolayer H-VSe$_2$. Both of these measured values are in excellent agreement with our DMC computed lattice constants. Additionally, we determined the near-equilibrium PES of both T and H 2D VSe$_2$ with DMC accuracy, which are both depicted in Fig. S6. 
 
The phase diagram presented in Fig. \ref{phase_diagram} is based on similar fits to data, where the $z$ displacement has been remapped to $d^{\textrm{V}-\textrm{Se}}$. This DMC phase diagram can directly be compared to the energy vs. strain DFT benchmarking calculations in Fig. S3, which emphasizes the need for an accurate representation of the phase boundary between the two phases. The freestanding geometries of both T and H lie in the energetic H phase, but a slice of the phase diagram along $d^{\textrm{V}-\textrm{Se}} = 2.505$ \AA\space indicates that the T phase becomes favorable over H at biaxial strain of $a \gtrsim 3.5$ \AA. This implies that in freestanding form, once T-VSe$_2$ is positively strained at least $\approx$ 2.5 $\%$, T phase is favored over H. Alternatively, if freestanding H-VSe$_2$ is positively strained at least $\approx$ 5 $\%$, T phase is also favored over H 
This strain can easily be accomplished by placing monolayer VSe$_2$ on a substrate with significant lattice mismatch. In fact, this type of mismatch has been reported to alter the material properties \cite{https://doi.org/10.1002/adma.201903779,vse2,doi:10.1063/6.0001402,10.3389/fmats.2021.710849}, significantly contributing to the controversies of T and H-VSe$_2$ (for energetic favorability, magnetic properties). Whether or not the changes in energetic favorability or magnetic properties with respect to the substrate are due to lattice mismatch or more complicated interactions between the substrate and the monolayer remains to be answered and is beyond the scope of this work, which has focused solely on the freestanding forms of T and H-VSe$_2$. However, such calculations can be employed for future work using higher order methods such as DMC. The proximity of the phase boundary between T and H phase (Fig. \ref{phase_diagram}) is emphasized by the small energy difference between the two phases (0.06(2) eV/f.u., at the equilibrium geometry) between the two curves. Since this energy difference is so close to room temperature ($\approx$ 0.024 eV), this implies that a process such as thermal annealing can easily induce a phase transition. In fact, recently it was demonstrated that a structural phase transition of multilayer VSe$_2$ from T to H occurs through annealing at 650 K, along with a metal-insulator transition \cite{struc-phase}.

To gain a deeper understanding of the magnetic properties of 2D T and H-VSe$_2$, we extracted the spin densities (using a trial wavefunction at $U$ = 2 eV and 24 atom supercell at the final equilibrium geometry predicted by DMC/line-search). The spin density isosurfaces of each phase ($\rho_{\textrm{up}}$ - $\rho_{\textrm{down}}$) are depicted in the insets of Fig. \ref{spindensity} a) and c) for T-VSe$_2$ and H-VSe$_2$ respectively. For both phases, we observe the V atoms are highly spin-polarized, while the Se atoms are slightly antiparallel with respect to the V atoms. For more calculation details regarding spin density, see SI.

We went on to plot the radial averaged spin densities as a function of distance, separately for V and Se for T and H-VSe$_2$ (depicted in Fig. \ref{spindensity} a) - d)). This allows us to view the spatial variations in spin density. Additionally, we benchmarked these V and Se radially averaged densities with PBE+$U$ ($U$ = 2 eV) using NC pseudopotentials at the equilibrium geometry (the calculation required to create the trial WF for the subsequent DMC runs). As seen in Fig. \ref{spindensity} a) and c), there is a substantial difference in the V spin density between DMC and PBE+$U$ ($U$ = 2 eV) for both T and H phase. This same substantial difference between DMC and PBE+$U$ also occurs for the total charge density. This discrepancy is most prevalent near the radial density peak (peak of $d$ orbital) and can be attributed to the fact that DFT functionals (even with the added Hubbard correction) tend to delocalize and unsuccessfully capture 3$d$ orbitals. This large discrepancy in the spin densities highlights the need for more accurate, many-body computational methodologies for correlated materials such as VSe$_2$, where DFT fails. In contrast, there is closer agreement between the DMC and PBE+$U$ spin densities for Se in T and H-VSe$_2$ (see Fig. \ref{spindensity} b) and d). 

Finally, we estimated the site-averaged atomic magnetic moments per V and Se for both T and H phase by integrating the DMC and PBE+$U$ spin densities depicted in Fig. \ref{spindensity}. At the DMC level, we estimated a magnetic moment of 1.06(2) $\mu_\textrm{B}$ for V and -0.09(2) $\mu_\textrm{B}$ for Se in T-VSe$_2$ and a magnetic moment of 1.02(1) $\mu_\textrm{B}$ for V and -0.14(1) $\mu_\textrm{B}$ for Se in H-VSe$_2$. At the PBE+$U$ ($U$ = 2 eV) level, we estimated a magnetic moment of 1.30 $\mu_\textrm{B}$ for V and -0.12 $\mu_\textrm{B}$ for Se in T-VSe$_2$ and a magnetic moment of 1.40 $\mu_\textrm{B}$ for V and -0.15 $\mu_\textrm{B}$ for Se in H-VSe$_2$. Consistent with the radial spin density results in Fig. \ref{spindensity}, we find that the DMC and PBE+$U$ magnetic moments for Se are in much closer agreement than for V (for both T and H phase). By analyzing the spin densities and obtaining the on-site magnetic moments, we obtain a clear picture of how the magnetization of each ion depends on the computational method used, serving as a benchmark for the magnetic properties of 2D VSe$_2$.

In this work, we used a combination of DFT, DMC and a recently developed surrogate Hessian line-search optimization technique to resolve the previously reported discrepancy in structural parameters and relative phase stability of monolayer T-VSe$_2$ and H-VSe$_2$. Using 
 these methods, we determined the lattice constant and V-Se distance (with DMC accuracy) to be 3.414(12) \AA\space and 2.505(7) \AA\space respectively for T-VSe$_2$ and 3.335(8) \AA\space and 2.503(5) respectively for H-VSe$_2$. In addition, we find the relative energy between the phases (T - H) to be 0.06(2) eV/f.u. at the DMC level, indicating that in freestanding form, H-VSe$_2$ is more energetically favorable than T-VSe$_2$. We went on to obtain a phase diagram between T and H phase from the PES and determined that a phase transition can be induced by strain or mechanisms such as thermal annealing. Additionally, we benchmarked the magnetic properties such as spin density and on-site magnetic moment for both phases and find substantial differences between DMC and DFT. The results of this study demonstrate the successes of the DMC method coupled with the surrogate Hessian line-search structural optimization technique when applied to a 2D magnetic system. The estimates for lattice constant, bond distance, relative phase energy and the extracted structural-dependent phase diagram assist in clarifying previously inconclusive theoretical and experimental results regarding T and H phase VSe$_2$.

 \section{Code Availability Statement}
Software packages mentioned in the article can be found at https://github.com/usnistgov/jarvis. Please note that the use of commercial software (VASP) does not imply recommendation by the National Institute of Standards and Technology.

 \section{Competing interests}
The authors declare no competing interests.

\section{acknowledgments}
The authors thank the National Institute of Standards and Technology for funding, computational, and data-management resources. The authors thank Dr. Kamal Choudhary and Dr. Francesca Tavazza for fruitful discussions. We acknowledge grants of computer capacity from the Finnish Grid and Cloud Infrastructure (persistent identifier  urn:nbn:fi:research-infras-2016072533).

\section*{References}
\bibliography{Main}

\end{document}


\maketitle
\section{\label{sec:results}Computational Methods}

Density functional theory (DFT) benchmarks for the T and H phase of 2D VSe$_2$ were performed using the Vienna Ab initio Simulation Package (VASP) code with projector augmented wave (PAW) pseudopotentials \cite{PhysRevB.54.11169,PhysRevB.59.1758}. For these calculations, the local density approximation (LDA)\cite{PhysRev.136.B864}, Perdew-Burke-Ernzerhof (PBE)\cite{PhysRevLett.77.3865}, and strongly constrained and appropriately normed (SCAN)\cite{PhysRevLett.115.036402} meta-GGA functionals were used with the added Hubbard correction ($U$) \cite{PhysRevB.57.1505} to treat the on-site Coulomb interaction of the $3d$ orbitals of the V atoms. At least 20 \AA\space of vacuum was given between periodic layers of VSe$_2$ in the $c$-direction. In addition, we used a reciprocal grid of 24x24x1 and a kinetic energy cutoff of 400 eV.

Our Quantum Monte Carlo (QMC) simulations used DFT-PBE to generate the trial wavefunction for fixed-node diffusion Monte Carlo (DMC) calculations. The Quantum Espresso (QE) \cite{Giannozzi_2009} code was used for our DFT calculations to create the trial wavefunction. This trial wavefunction was created for the ferromagnetic configuration of 2D VSe$_2$ using different $U$ values with the goal of variationally determining the optimal nodal surface ($U$ value that yields the lowest total energy). For V, we used norm-conserving (NC) RRKJ (OPT) pseudopotentials \cite{PhysRevB.93.075143} and for Se, we used NC Burkatzki-Fillipi-Dolg (BFD) pseudopotentials \cite{doi:10.1063/1.2741534}. After testing at the DFT level, a kinetic energy cutoff of 4,080 eV (300 Ry) and a k-grid of 6x6x1 was used (see Fig. S1 and S2) to generate trial wavefunctions for DMC. To accelerate the line-search method convergence for the metallic T phase, we increased the k-grid to 12x12x1. 

After the trial wavefunction was generated with DFT, Variational Monte Carlo (VMC) and DMC \cite{RevModPhys.73.33,Needs_2009} calculations were performed using the QMCPACK \cite{Kim_2018,doi:10.1063/5.0004860} code. The single determinant DFT wavefunction is converted into a many-body wavefunction by use of the Jastrow parameters \cite{PhysRev.34.1293,PhysRev.98.1479}, which assist in modeling electron correlation with the goal of reducing the statistical uncertainty in DMC calculations \cite{PhysRevLett.94.150201,doi:10.1063/1.460849}. Up to two-body Jastrow \cite{PhysRevB.70.235119} correlation functions were included, where the linear method \cite{PhysRevLett.98.110201} was used to minimize the variance and energy of the VMC energies. The cost function of the variance optimization is 100 $\%$ variance minimization and the cost function of the energy optimization is split as 95 $\%$ energy minimization and 5 $\%$ variance minimization, which has been proven to reduce the uncertainty of DMC calculated results \cite{PhysRevLett.94.150201}. The Nexus \cite{nexus} software suite was used to automate the DFT-VMC-DMC workflow. The locality approximation \cite{doi:10.1063/1.460849} was used to evaluate the nonlocal part of the pseudopotentials in DMC and an optimal timestep of 0.01 Ha$^{-1}$ was determined for DMC simulations due to the fact that it yielded an acceptance ratio greater than 99 $\%$ (see Table S1). A full summary of the VMC and DMC methods can be found in reference \cite{RevModPhys.73.33}.

The total charge density and spin density was extracted from our DMC calculations. The spin density is defined as the difference between the spin-up contribution to the total charge density and the spin-down contribution to the total charge density ($\rho_{up}-\rho_{down}$). We used an extrapolation scheme on the DMC charge densities with the goal of eliminating the bias that occurs from using a mixed estimator. Since the charge density estimator does not commute with the fixed-node Hamiltonian, the DMC charge density was obtained from a mixed estimator between the pure fixed-node DMC and VMC densities. The extrapolation formula takes the form\cite{RevModPhys.73.33}:

\begin{equation} \label{rho1}
\rho_1 =2\rho_{\textrm{DMC}}-\rho_{\textrm{VMC}}+\mathcal{O}[(\Phi-\Psi_{\textrm{T}})^2]
\end{equation}
where $\rho_{\textrm{DMC}}$ and $\rho_{\textrm{VMC}}$ are the DMC and VMC charge densities respectively. $\Phi$ is the trial wavefunction from the DMC Hamiltonian and $\Psi_{\textrm{T}}$ is the trial wavefunction from VMC.

In addition, we integrated the DFT+$U$ and DMC spin densities up to a cutoff radius $r_{cut}$ (which we define as 1.34 \AA\space, due to the fact that it is approximately half of the V-Se bond distance in 2D T and H-VSe$_2$) in order to estimate the site-averaged atomic magnetic moment per V and Se. To obtain these magnetic moments per atom ($M_A$), we sum over the spherically interpolated spin densities: 

\begin{equation} \label{M_a}
M_A = 4\pi \int_0^{r_{cut}}r^2 \rho_s(r)dr \approx 4\pi \sum_{i=0}^{r_{cut}/\Delta r}r_i^2 \rho_s(r_i)\Delta r
\end{equation}
where $r_i$ is the distance from the center of the atom to a given point on the grid and $\Delta r$ is the radial grid size.

To optimize the structural parameters of both T and H-VSe$_2$ according to the DMC potential energy surface (PES), we use a surrogate Hessian accelerated optimization method \cite{doi:10.1063/5.0079046}. In the method, we consider the PES around equilibrium as the second-order expansion in Wyckoff parameter space, $p$:
\begin{equation} \label{ls1}
E(p)=E_0+\frac{1}{2}(p-p_0)^{T}H_p(p-p_0),
\end{equation}
where $H_p$ is the Hessian, or the force-constant matrix, $E_0$ is the energy minimum and $p_0$ the energy-minimizing parameters. Diagonalizing the parameter Hessian, i.e., $H_p = U^T \Lambda U$, forms an optimal basis for a conjugate line-search in the parameter space, namely the eigenvectors $U$. The line-searches along $U$ can be conducted in parallel, and ideally, they locate the minimum in just one parallel iteration within the quadratic region. Here, we conduct the line-search according to a set of 2 parameters: the lattice constant $a$ and the Wyckoff parameter $z$, which is the unsigned displacement of the Se atoms along the $z$ axis (see Fig. 1). For reporting purposes, the line-search parameters $a$ and $z$ are remapped to $a$ and $d$, where $d$ is the V-Se distance.

In the surrogate Hessian scheme, we obtain a cheap but relatively accurate Hessian from DFT, and use it to the inform line-search on the DMC PES, in particular by providing the search directions. We also resample the DFT PES to predict fitting errors. Thus, we may minimize the computational cost of the DMC runs, while maintaining an error tolerance. The surrogate DFT PES was based on QE with a 4,080 eV (300 Ry) cutoff using PBE with no DFT+$U$ correction. The DMC PES was based on DFT-PBE with $U$ = 2 eV orbitals and finite-size extrapolation through supercell sizes of 9 and 24 atoms. Each line-search was based on a 3rd order polynomial fit and set to contain 7 points, or displaced geometries, totaling 13 energy evaluations per phase, per iteration. However, alternative techniques, including (bi)polynomial fitting, were used in some parts to incorporate auxiliary DMC data and ensure convergence to the quadratic region. Effectively, two parallel line-search iterations for both phases were carried out, and the convergence was claimed in the absence of significant displacements.

\begin{figure}
\begin{center}
\includegraphics[width=14cm]{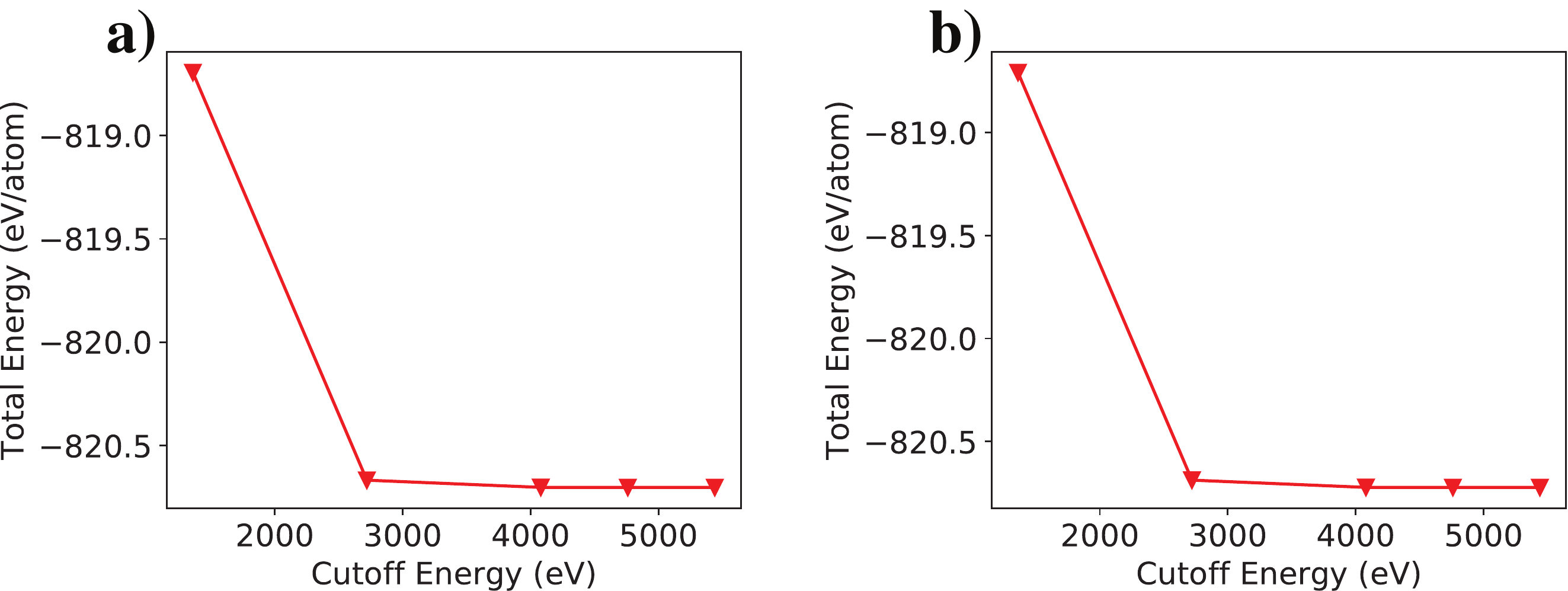}
\caption{The total energy per atom of the unit cell (3 atoms) of 2D a) T-VSe$_2$ and b) H-VSe$_2$ as a function of plane wave cutoff energy
for the norm-conserving pseudopotentials calculated with DFT using the PBE functional at a k-point grid of 6x6x1.
The results show a converged cutoff energy of 4,080 eV (300 Ry) for both phases.}
\label{cutoff}
\end{center}
\end{figure}

\begin{figure}
\begin{center}
\includegraphics[width=14cm]{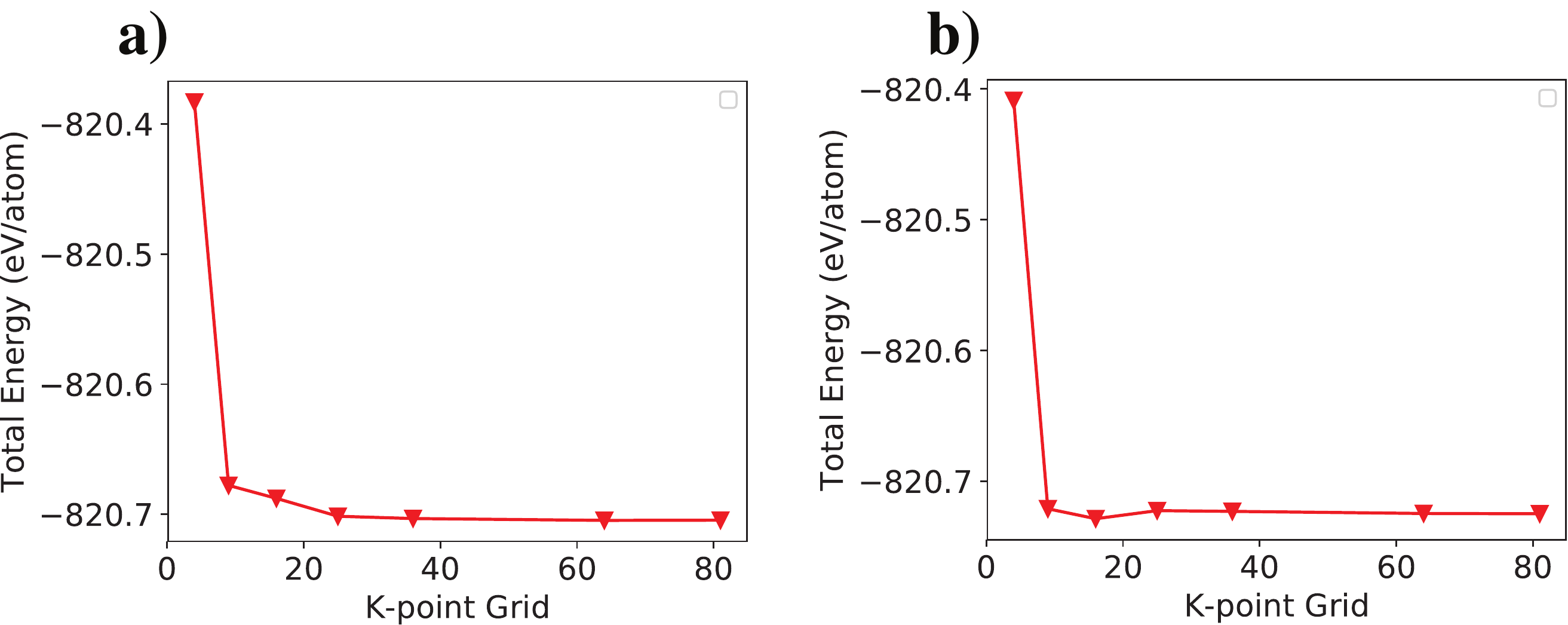}
\caption{The total energy per atom of the unit cell (3 atoms) of 2D a) T-VSe$_2$ and b) H-VSe$_2$ as a function of K-point grid
for the norm-conserving pseudopotentials calculated with DFT (PBE) at the converged cutoff energy (see Fig. S1).
The results show a converged k-point grid of 6x6x1 (36) for both monolayers. The number of K-points was scaled appropriately to obtain the converged grid depending on the supercell size and shape for all DFT and DMC calculations.}
\label{kpoint}
\end{center}
\end{figure}

\begin{table}[]
\caption{
Tabulated results for the DMC timestep convergence of a 12 atom cell of 2D T-VSe$_2$ and H-VSe$_2$. The acceptance ratio of 0.99 indicates that 0.01 Ha$^{-1}$ is an appropriate timestep to use for all subsequent DMC simulations.   }
\begin{tabular}{l|l|l|l}
\hline
T-VSe$_2$   &                &           &                 \\
\hline
\hline
Timestep (Ha$^{-1}$) & DMC Total Energy (Ha) & Error (Ha) & Acceptance Ratio \\
\hline
0.02                         & -361.730                            & 0.001                          & 0.985                                \\
0.01                         & -361.709                            & 0.002                          & 0.994                                \\
0.005                        & -361.709                            & 0.003                          & 0.997                                \\
0.002                        & -361.702                            & 0.002                          & 0.999                                \\
\hline
\hline
H-VSe$_2$   &                 &           &                \\
\hline
\hline
Timestep (Ha$^{-1}$) & DMC Total Energy (Ha) & Error (Ha) & Acceptance Ratio \\
\hline
0.02                         & -361.673                            & 0.001                          & 0.985                                \\
0.01                         & -361.657                            & 0.002                          & 0.994                                \\
0.005                        & -361.654                            & 0.002                          & 0.998                                \\
0.002                        & -361.657                            & 0.003                          & 0.999                     \\
\hline
\end{tabular}
\end{table}

\begin{figure*}
\begin{center}
\includegraphics[width=15cm]{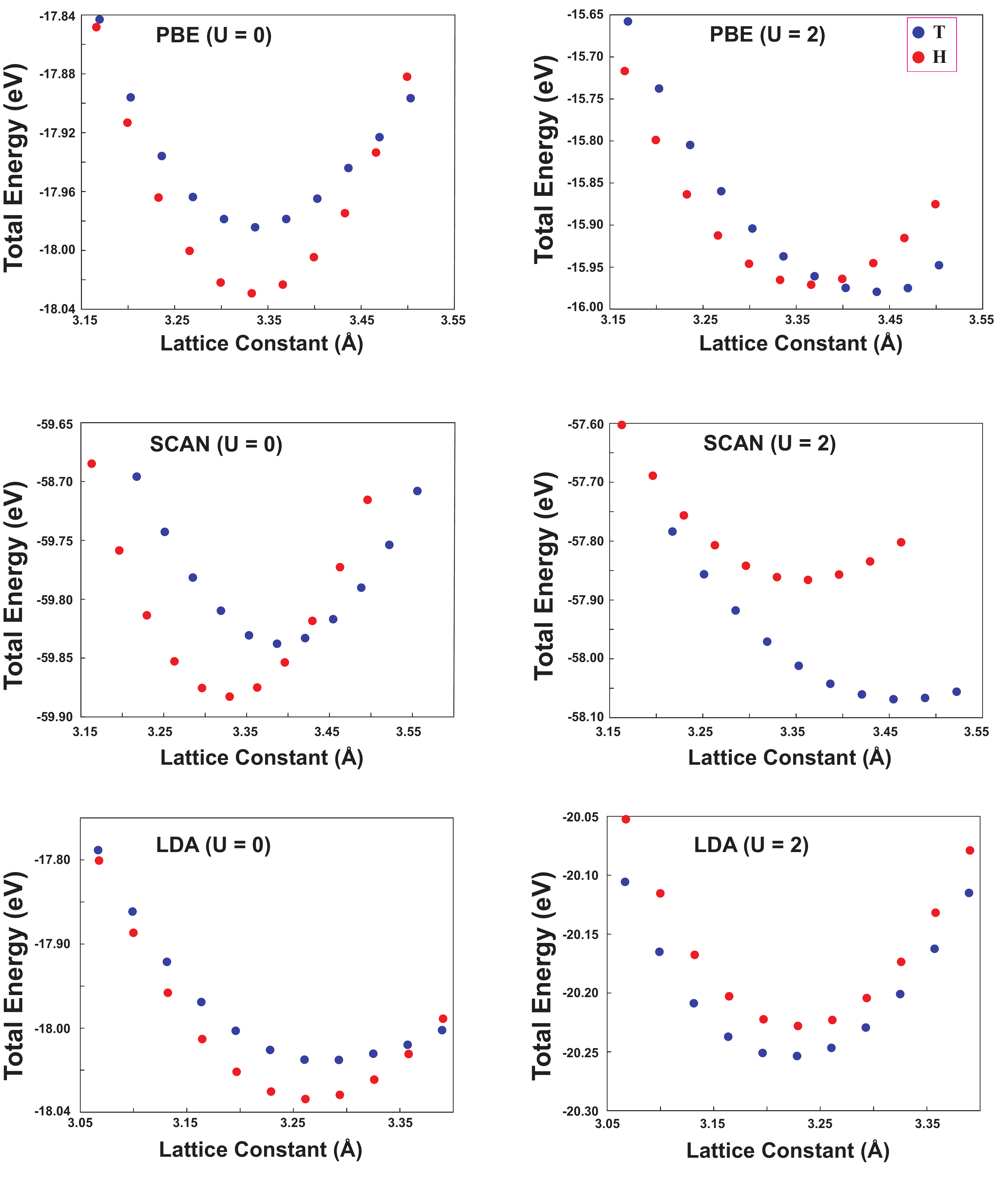}
\caption{Total energy as a function of lattice strain for T (blue) and H (red) phase 2D VSe$_2$, calculated with various functionals and $U$ values. Density functionals include LDA, PBE, and SCAN.   }
\label{dft-bench}
\end{center}
\end{figure*}

\begin{figure}
\begin{center}
\includegraphics[width=8cm]{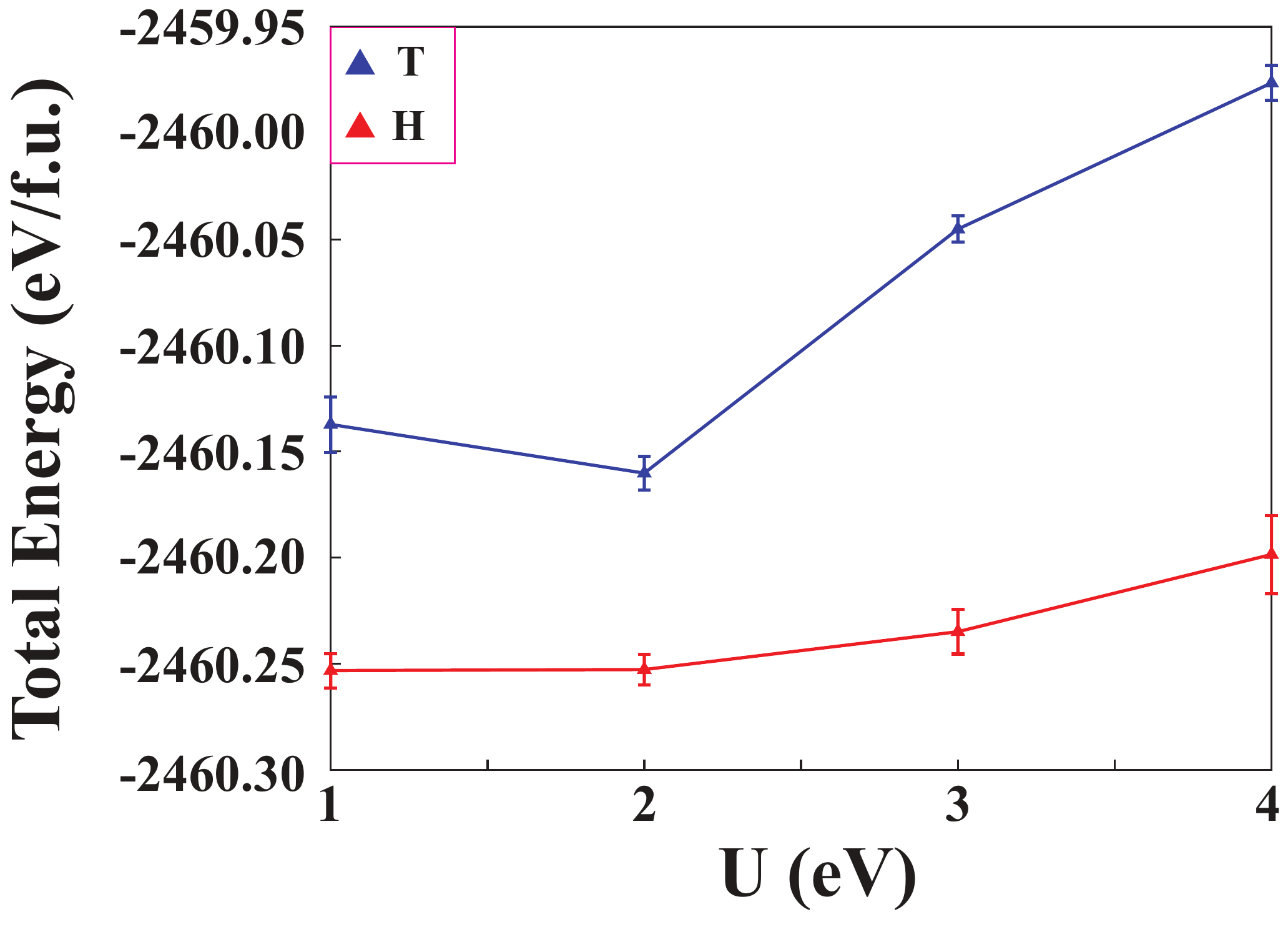}
\caption{DMC calculated total energies of a 24-atom supercell (normalized per formula unit (f.u.)) of 2D T (blue) and H (red) phase VSe$_2$ calculated as a function of the $U$ parameter used to variationally determine the optimal trial wave function. The DMC error bars represent the standard error about the mean.}
\label{u-tune}
\end{center}
\end{figure}

\begin{figure}
\begin{center}
\includegraphics[width=8cm]{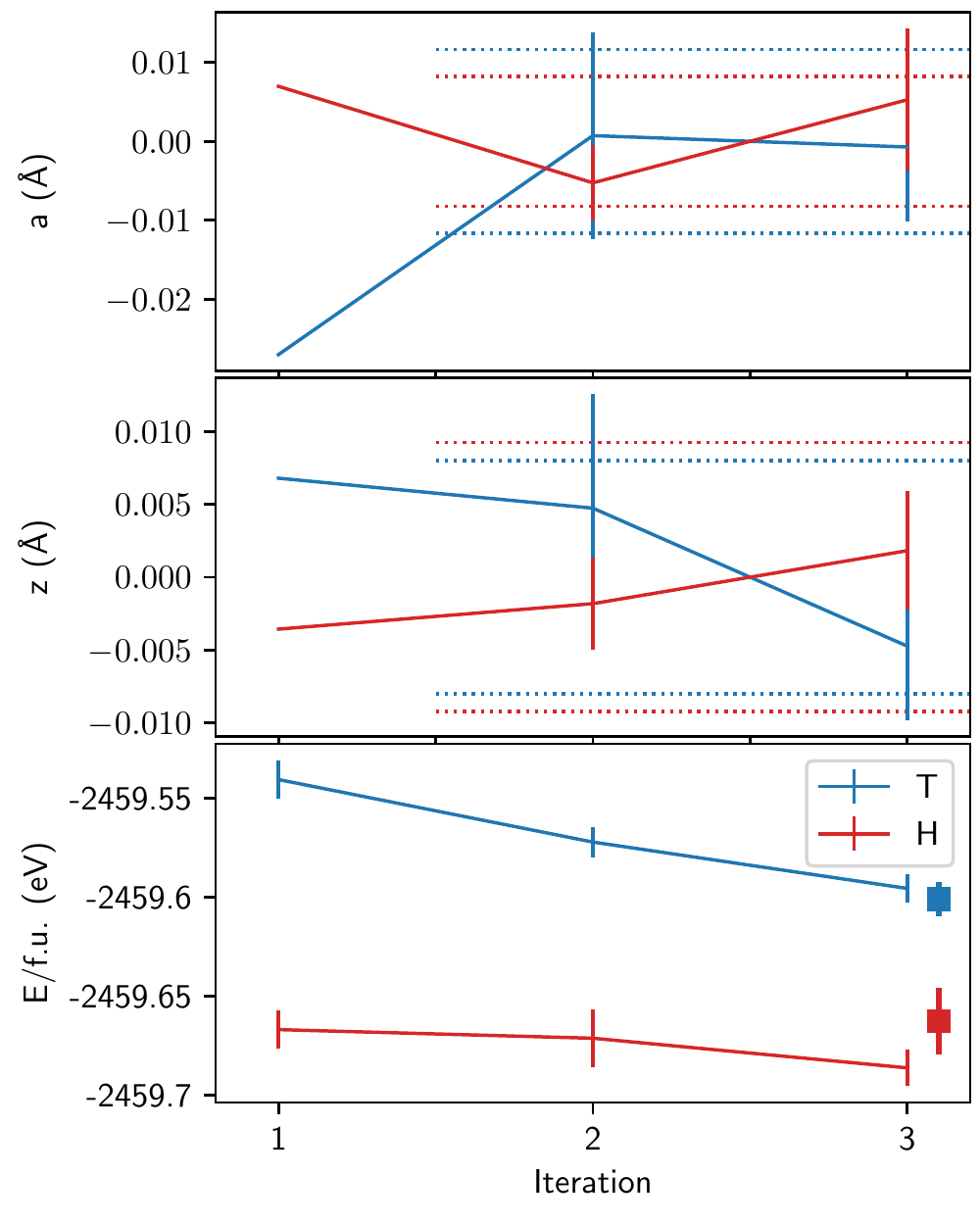}
\caption{The convergence of the $a$ and $z$ parameters and DMC energies per f.u. for both T (blue) and H (red) phase of 2D VSe$_2$ based on parallel line-search iterations along the DMC PES. The starting parameters (iteration 1) are from DFT, the zero offset is the mean over iterations 2 and 3, and dotted lines indicate the error tolerances for each case (95 \% confidence). The DMC energies from respective equilibrium geometries are plotted with 1SEM (one standard error of the mean) uncertainties, with extra squares marking energies from the predicted minimum geometry.}
\label{ls_converge}
\end{center}
\end{figure}

\begin{figure*}
\begin{center}
\includegraphics[width=14cm]{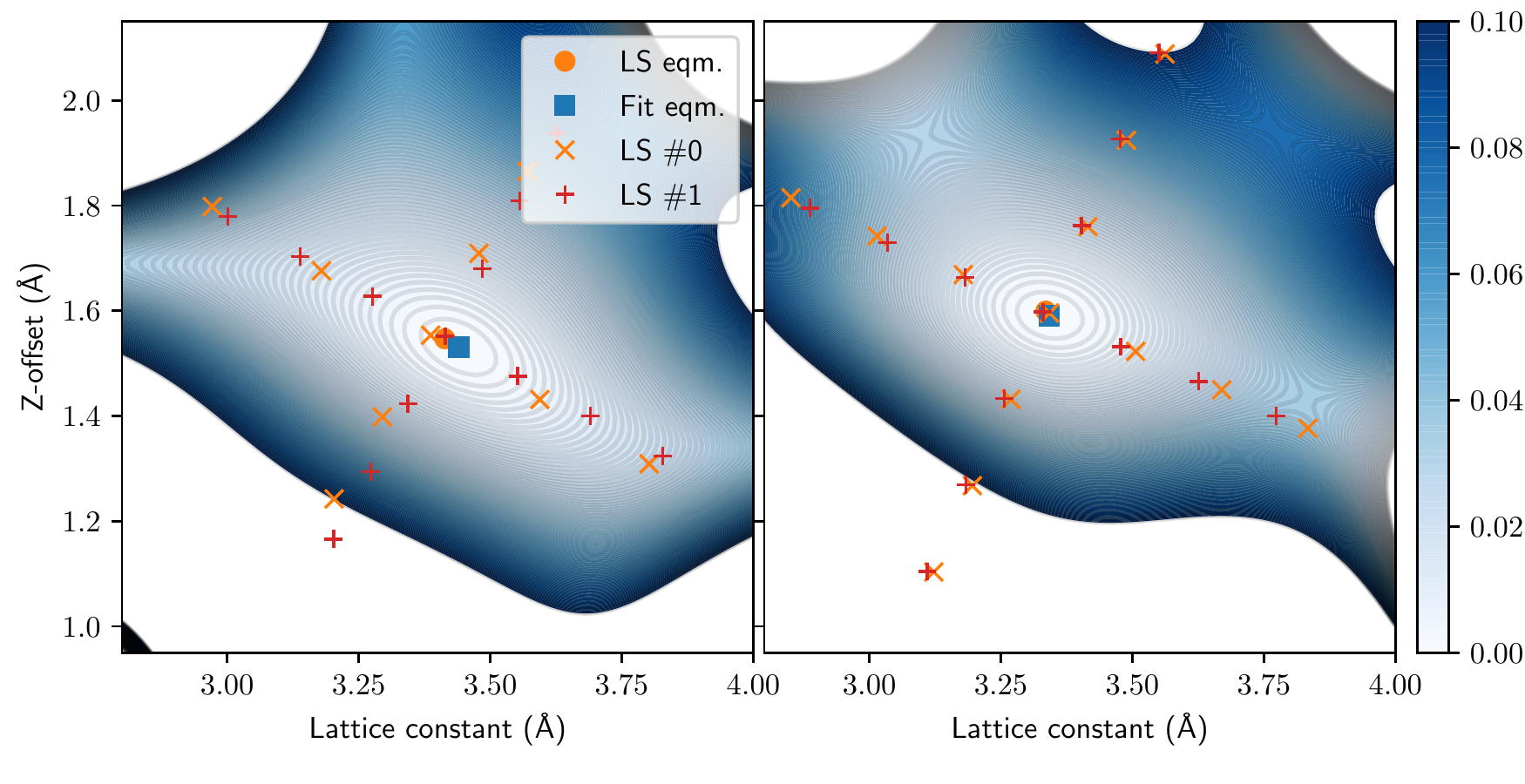}
\caption{Contour reconstructions of the DMC PESs (eV) of T (left) and H (right) phases of 2D VSe$_2$ with respect to $a$ and $z$ parameters. The contours are based on bicubic fits to sparse data, and thus, subject to biases and statistical uncertainties not indicated in the figures. The markers ('x' and '+') indicate data points from two parallel line-search iterations.}
\label{search}
\end{center}
\end{figure*}

\newpage
\bibliography{si}